# Study on the Assessment of the Quality of Experience of Streaming Video

Aleksandr Ivchenko
Laboratory of Multimedia systems and technology
Moscow Institute of Physics and Technology
Moscow, Russia
ivchenko.a.v@phystech.edu

Pavel Kononyuk
Laboratory of Multimedia systems and technology
Moscow Institute of Physics and Technology
Moscow, Russia
kononyuk@phystech.edu

Alexander Dvorkovich
Laboratory of Multimedia systems and technology
Moscow Institute of Physics and Technology
Moscow, Russia
dvorkovich.av@mipt.ru

Liubov Antiufrieva
Laboratory of Multimedia systems and technology
Moscow Institute of Physics and Technology
|Moscow, Russia
antyufrieva@mipt.ru

*Abstract*—Dynamic adaptive streaming over HTTP provides the work of most multimedia services, however, the nature of this technology further complicates the assessment of the QoE (Quality of Experience). In this paper, the influence of various objective factors on the subjective estimation of the QoE of streaming video is studied. The paper presents standard and handcrafted features, shows their correlation and p-Value of significance. VQA (Video Quality Assessment) models based on regression and gradient boosting with SRCC reaching up to 0.9647 on the validation subsample are proposed. The proposed regression models are adapted for applied applications (both with and without a reference video); the Gradient Boosting Regressor model is perspective for further improvement of the quality estimation model. We take SQoE-III database, so far the largest and most realistic of its kind. The VQA (video quality assessment) models are available at https://github.com/AleksandrIvchenko/QoE-assesment

## I. Introduction

Recently there was an explosive growth in multimedia services. Most video content is delivered through Dynamic Adaptive Streaming over HTTP (DASH) technologies [1]. For example, this is how the YouTube and Netflix services works using TCP [2]–[3] as underlying transport protocol.

Unlike UDP-based [4]–[5] data transfer, where video and audio packets are allowed to be lost and playback is performed using the existed data and damaged data decoder strategies, TCP guarantees data delivery by means of channel control and retransmissions of lost and damaged packets. However, this leads to other problems [48]: with insufficient bandwidth (below the video transmission speed) or/and packets losses, the playback buffer is depleted and stalling events occur. In addition, there is congestion control. "TCP window" determines the number of packets (typical values are from 1 to 10) that are sent simultaneously, after which there is a waiting phase for confirmation (ACK packet). During the session, the window gradually increases. Congestion control, noticing packet loss, thinks that the network is overloaded, that there are too many packets in network, and reduces the window (usually 2 times).

To work in a dynamically changing channel, video is encoded at various bitrate levels – quality levels. The server provides the transmission of several video streams transmitted in small segments via HTTP, data about available quality levels and the location of the segments are transmitted in the manifest. The decoder on the user side provides adaptation of the playback speed.

On the user's side, quality levels are switched to match the current bandwidth, buffer status, and selected playback speed. The ABR (adaptive bitrate streaming) algorithm is responsible for choosing the next segment [6]. In this case, the optimization metric should be maximization of the user's QoE.

This leads tree typical impairments for DASH video steaming: delay in initial buffering, stalling events (video freezing) and quality switchings. Besides, content should be available both on stationary multimedia systems with widescreen screens, and on small-format mobile devices. In this study we focus only on high size screens.

The common way of QoE measuring is the creation of expert groups that, under certain conditions of the experiment, give a rating on a rank scale, then these estimates are averaged (as a rule with some pre-processing – cleaning from emissions), thus obtaining a MOS (Mean Opinion Score). However, nowadays using this approach is almost impossible, because of the amount of video. In addition, there is an effect of learning: the expert's estimation begins to differ significantly from the average user. There is no single, most effective methodology for an objective estimation of the subjective QoE.

In this paper, we develop the research of the Zhengfang Duanmu group [7], with the SqoE-III database, the most extensive one now. The main problem of previous studies was an insufficient database for the experiment, both in the number of video sequences and in the experiment itself: the presence of initial buffering, delay events, and quality level switching. In addition, previous experiments were not consistent with the actual behavior of the network, and the introduction of distortion was artificial. SqoE-III has some drawbacks. However, its analysis allows us to develop the conditions for further experiments, including the correct expansion of the database.

In this paper, we analyze the influence of classic and handcrafted metrics on QoE, investigate the dependence of the MOS curve on the integral indicator of objective quality and propose several variants of objective assessment models for VQA.

Section II provides a small overview of the current state of art.

In the Section III brief overview of SQoE-III is given. The section contains an analysis of the contents of the database, a description of the metrics (attributes) and their impact on MOS.

Section IV contains experiments: the study of the shape of the curve dependence, the construction of VAQ models based on the gradient boosting and linear models.

Section V contains an interpretation of the results obtained, the importance of the characteristics of the models and a description of the principle of operation..

Section VI contains a comparison of the quality of the obtained models with world practice and summarizes the results of the work.

The Discussion section describes the shortcomings and options for further research.

## II. RELATED WORK

There are quite a lot of metrics with which you can evaluate the quality of the video. Many of them were created for specific experimental conditions, sub-specific multimedia systems, and they were qualitatively evaluated on different datasets. In 2004, ITU conducted large-scale tests that showed the statistical equivalence of the most effective metrics at that time, both among themselves and in comparison with the simplest PSNR [8].

The classification of metrics is presented below and several works close to our work are given.

According to the information that the VQA uses, the classification proposed by Takahashi et al.[9] can be used.as shown in Figure 1.

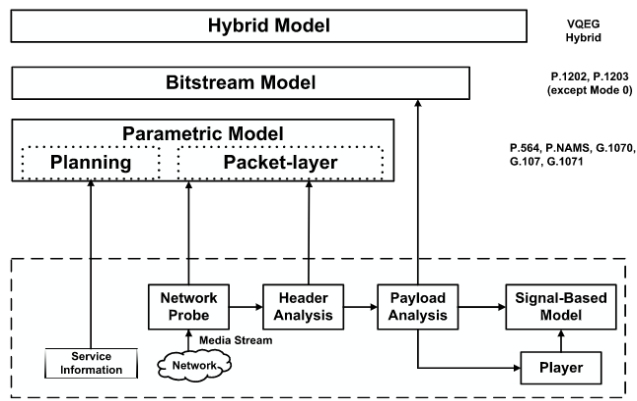

Fig. 1. VQA models types. The proposed methods can be classified as Bitsteram model

Signal-based models use only decoded signal to estimate quality. They are based on an estimate of the distortion of the signal passing through an unknown system and has three types: Full Reference (FR), Reduced Reference (RR) and No Reference (FR).

- FR models requires all information about input signal (reference video) and computes frame-by-frame comparison. This class includes MSE, PSNR, HVS-PSNR SSIM, MS-SSIM [10] and ITU-T Recommendations [11]–[13].

- RR requires part of the input system information. This class includes ITU-T J.246 Recommendation [14], models of Z. Wang [15]–[16] and others. For example, such metrics need knowledge about types of distortions, content type (natural scenes or computer graphics).

- NR modes do not require any system's input information and based only on received signal. This class includes DIIVINE, BRISQUE, BLIINDS and NIQE [17]–[20].

Parametric models that bases on network layer can be called QoS-metrics. This class divided on Packet-layer models and Planning models. First uses actual information from packet headers like delay, jitter, packet loss and other. It has no information about video playback. This class includes ITU-T Rec P.1201 [21]. Planning models uses exactly the same information, but taken from planning phase on user's side ((ITU-T Rec. G.1070 [22], E-model ITU-T Rec. G.107 [23]).

Bitstream models take into account the encoded bit streamed packet layer information. Features such as bitrate, framerate, Quantization Parameter (QP), PLR, motion vector, macroblock size (MBS), DCT coefficients, etc. are extracted and used as input to the model. ITU-T Rec. P.1202, with ITU-T Rec. P.1203 being the most recently approved recommendation [24].

Hybrid models combine data from the three previous models and therefore are the most effective, but more complex in practical application.

The authors of the article [25] proposed an interesting solution when working with the YouTube service. They analyzed traffic and metrics. Back-propagation neural network and random forest were applied to evaluate stalling and estimate initial buffering delay.

The ITU-T Recommendation series P.1203 proposes a parametric bitstream-based model for the quality assessment is close to our work, using ensemble of trees and similar features [24].

Model P.1203 is refined by Zaixin Yang et al. [26]. The buffering inputs required for the model is too complex and redundant, and it does not consider the impact of the worst video quality on QoE.

A significant contribution was made by Ivan Bartolec et al. describing the impact of user interaction scenarios with a multimedia service using YouTube as an example. This paper explores the effects of playback-related interactions, such as video skipping, pausing, and seeking [27].

A refinement to the construction of machine learning models is work. In it, the authors use the evaluation model with memory – the LSTM neural network including impacts of quality variations, stalling events, and content characteristics (spatial and temporal complexity). Previous experience is one of the factors influencing assessment and a neural network with memory can evaluate it [28].

## III. DATABASE ANALYSIS

The SqoE-III dataset consists of 20 reference videos in a resolution of 1920x1080 for 10 seconds, consisting of various types of content: people, animals, nature scenes, architecture, screen recordings and computer-generated videos (Table I). Videos are presented with various spatial and temporal information (SI and TI features).

TABLE I. DESCRIPTION OF REFERENCE VIDEOS

| Name | FPS | SI | TI | Description |
|---|---|---|---|---|
| BigBuckBunny | 30 | 96 | 97 | Animation, high motion |
| BirdOfPrey | 30 | 44 | 68 | Natural scene, smooth motion |
| Cheetah | 25 | 64 | 37 | Animal, camera motion |
| CostaRica | 25 | 45 | 52 | Natural scene, smooth motion |
| CSGO | 60 | 70 | 52 | Game, average motion |
| FCB | 30 | 80 | 46 | Sport, average motion |
| FrozenBanff | 24 | 100 | 88 | Natural scene, smooth motion |
| Mtv | 25 | 112 | 144 | Human, average motion |
| PuppiesBath | 24 | 35 | 45 | Animal, smooth motion |
| RoastDuck | 30 | 60 | 84 | Food, smooth motion |
| RushHour | 30 | 52 | 20 | Human, average motion |
| Ski | 30 | 61 | 82 | Sport, high motion |
| SlideEditing | 25 | 160 | 86 | Screen content, smooth motion |
| TallBuildings | 30 | 81 | 13 | Architecture, static |
| TearsOfSteel1 | 24 | 53 | 66 | Movie, smooth motion |
| TearsOfSteel2 | 24 | 56 | 11 | Movie, static |
| TrafficAndBuilding | 30 | 66 | 15 | Architecture, static |
| Transformer | 24 | 72 | 56 | Movie, average motion |
| Valentines | 24 | 40 | 52 | Human, average motion |
| ZapHighlight | 25 | 97 | 89 | Animation, high motion |

The stand for experiments was created, that emulates the streaming service, including a bandwidth shaper and a delay emulator (Linux traffic control utility). Logging was performed on the client side, including bit rate, initial buffering duration, timestamps of the beginning and end of delay events. According to the log, the sessions were reconstructed by concatenating the "streaming bit rate representation" with the addition of blank frames to the test video to simulate initial buffering and inserting identical frames at the time of buffering to simulate a stalling event. The loading indicator (both for initial buffering and for stopping) was implemented as a spinning wheel. Each video was encoded in 11 variants by the x264 encoder to cover various quality levels. The videos are divided into segments for 2 seconds; the characteristics of the system can change inside the segment. The dataset is described in details in the original work [7].

### A. Feature engineering

The original dataset, which is a MATLAB model, iteratively describing each object, was transformed into a table for the convenience of researchers. It also added a number of new features that will be discussed below.

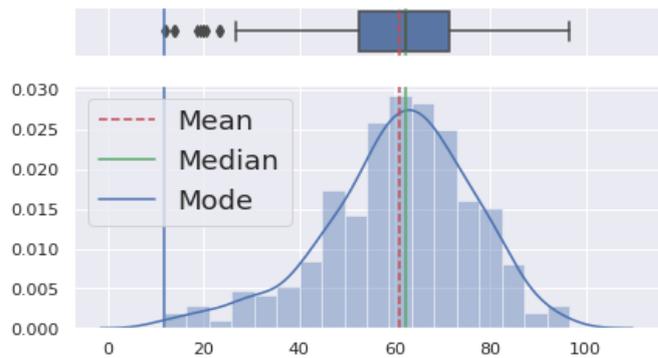

Fig. 2. MOS scores didtribution

Figure 2 shows the distribution of the MOS. The distribution is normal, data ranging from 11.64 to 96.63. The scales ranked, fractional values are obtained after averaging and are new rank estimates.

To begin with, it was decided to use the standard client-side QoE media metrics and check their P-value of significance [29]. Since we are calculating the correlation of the absolute and rank scales, we use a coefficient suitable for a lower level (rank) scale – Spearman's correlation coefficient.

Standard quality metrics (Table II):

1. Initial buffer time – time from the start of playback initialization to the start of video rendering on the user side.

2. Rebuffer percentage – ratio of the total duration of stalling events to the total playback time (excluding initial buffering).

$$\frac{\sum_i stalling\_event\_duration\_i}{duration - initial\_buffering\_time}$$

3. Rebuffer count – the number of rebuffering events.

4. Average rendered bitrate, kbps – the average bitrate (taking into account the duration of each bitrate level, this is discrete values from 235 till 7000 kbps).

5. Bitrate switch count - the number of bitrate switches.

6. Average bitrate switch magnitude, kbps – ratio of the sum of the modules of the pairwise bitrate difference to the number of bitrate switches.

$$\frac{\sum_i |bitrate_i - bitrate_{i-1}|}{\#bitrate\_switches}$$

7. Ratio on highest video quality level – ratio of the playback duration of the maximum possible bitrate to the playback duration (excluding initial buffering).

$$\frac{\sum_i \max\_playback\_quality\_duration_i}{duration - initian\_buffering\_time}$$

TABLE II. STANDARD QUALITY METRICS

| Quality metric | SRCC | P-value |
|---|---|---|
| Initial buffer time, seconds | -0,0303 | 0.52098 |
| Rebuffer percentage | -0,2733 | **3.8e-09** |
| Rebuffer count | -0,2505 | **7.21e-08** |

| Quality metric | SRCC | P-value |
|---|---|---|
| Average rendered bitrate, kbps | **0,5118** | **2.1e-31** |
| Bitrate switch count | *-0.1583* | **0.00075** |
| Average bitrate switch magnitude, kbps | *0,1082* | **0.02166** |
| Ratio on highest video quality level | 0,1172 | **0.01289** |

Values of Bitrate switch count and Average bitrate switch magnitude are different from original 0,1344 and 0,1583, that seems to be the misprint.

The null hypothesis means the absence of significant differences between the average values of the samples. P-level of significance – the probability of such or even more pronounced deviation from the average value, if the null hypothesis is true. The lower the value of P-value, the more reason to reject the null hypothesis. If the value is less than 0.05, we obtain a statistically significant deviation. Bold values can be considered statistically significant.

Together with the Average bitrate switch magnitude, the feature «Average *relative* bitrate switching magnitude» was introduced without a module, taking into account deterioration and improvement in quality.

Analyzing the Ratio on highest video quality level metric, it was noted that the most significant feature isn't the duration of the maximum possible bitrate, but the maximum in the session under consideration – Ratio on the highest sequence quality level:

$$\frac{\sum_i highest\_playback\_quality\_duration\_in\_sequense_i}{duration - initian\_buffering\_time}$$

Figure 3 shows SRCC and P-value change for a fraction of different quality levels.

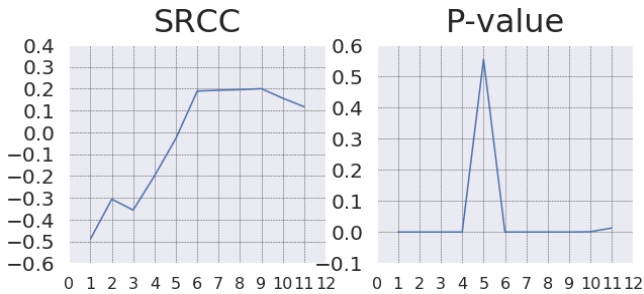

Fig. 3. Ratio on different video quality levels. Level 5 descrives medium vieo quality (640x480, 1050 kbps) and this ratio is not statistically significant.

The effect of the ratio on the minimum quality level is the most significant (SRCC -0.4895) so the feature Ratio on minimum sequence quality level was created. A slight decline after level 9 can be explained by the influence of other factors - at higher bitrates, more stalling events occur.

In addition, you can see that the effect of levels 6–9 is almost the same. If we introduce a feature that describes the duration of the playback of bitrates close to the maximum, you can get a good metric. Therefore, the feature "staying at levels 6–9" gives SRCC 0.4960 with a P-value of 2.6e-29, and the feature "staying at levels 6-11" equals SRCC 0.7586 and P-value 2.2e-85.

In addition, the feature of the constant bitrate during session was introduced. If you consider a subsample of sessions with such bit rates, then you get SRCC 0.8556 and 1.7e-16.

Besides, the following metrics (features) have been added (Table III):

1. Average video resolution – the average width multiplied by the average height.

2. Mean (seqPSNR) – average value of PSNR. Below you can find a variant of the applied reference-free formula without PSNR.

3. Investigation of separately lowering and increasing bitrate events:

- Bitrate_pos_changes_count – number of bitrate increase events.
- Bitrate_neg_changes_count – number of bitrate decrease events.
- Bitrate_max_pos_change – value of maximum bitrate increase.
- Bitrate_max_neg_change – value of maximum bitrate dercrease.
- Bitrate_mean_pos_change – value of average bitrate increase.
- Bitrate_mean_neg_change – value of average bitrate decrease.

4. Values of temporal and spatial information TI and SI. Their distribution in the common space is shown in Figure 4.

5. The frequency of occurrence of stalling events «Frequency of stalling» – the ratio of the number of stalling events to the total duration, including the initial delay.

6. Average duration of a stalling events, estimated in seconds.

7. Maximum duration of stalling, estimated in seconds – the longest stalling event duration in a session.

8. Frequency of switching – arequency of bitrate switching, ratio of the number of switches to the total duration.

TABLE III. ADVANTAGE NUMERICAL QUALITY METRICS

| Quality metric (features) | SRCC | P-value |
|---|---|---|
| Average *relative* Bitrate Switching magnitude | 0,1333 | **0.0046** |
| Ratio on highest sequence quality level | 0.1776 | **0.0002** |
| Ratio on minimum sequence quality level | **-0.4895** | **1.7e-28** |
| Ratio on sequence quality level max/2 | **0,7586** | **2,2e-85** |
| Average video resolution | **0.5497** | **6.9e-37** |
| Mean(seqPSNR) | **0,4606** | **5,2e-25** |
| Bitrate_pos_changes_count | 0.002 | 0.9660 |
| Bitrate_neg_changes_count | -0.1899 | **5e-5** |
| Bitrate_max_pos_change | 0.0531 | 0,2608 |
| Bitrate_max_neg_change | 0.1773 | **0,0002** |
| Bitrate_mean_pos_change | 0,0727 | 0,1236 |

| Quality metric (features) | SRCC | P-value |
|---|---|---|
| Bitrate_mean_neg_change | 0.1672 | **0,0004** |
| TI | 0.0697 | 0.1398 |
| SI | -0.0837 | 0.0762 |
| Frequency of stalling | -0.2142 | **4,5e-6** |
| Average duration of stalling events | -0.2615 | **1.8e-8** |
| Maximum duration of stalling | -0.2635 | **1.4e-8** |
| Frequency of switching | -0.726 | 0.124 |

It was studied what gives a better correlation with MOS: average or median values of every feature. There was a hypothesis that the median would be better, as more resistant to emissions, but everywhere a slightly better SRCC result showed an average value.

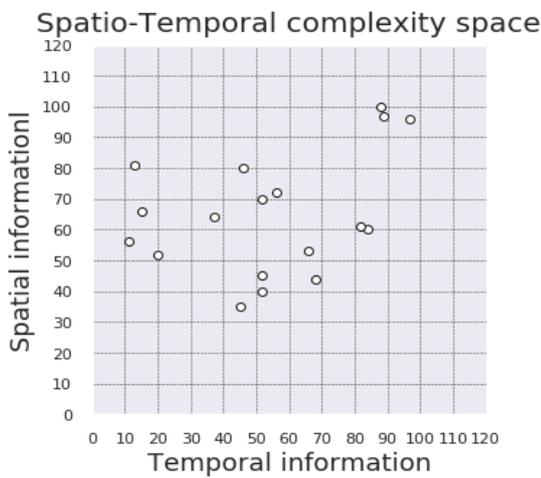

Fig. 4. Spatial perception information is based on the Sobel filter. This is the maximum value of the standard deviation from the Sobel function, applied sequentially to all frames (brightness density). Information on temporal perception is calculated as the maximum value of the standard deviation of the pixel difference (in the brightness plane) between consecutive frames [30].

### B. Motion and content type

At the Density-plot (Figure 5) for the amount of movement estimated in the dataset, there are smooth, high and camera motion differ slightly, static and average motion differ noticeably from them.

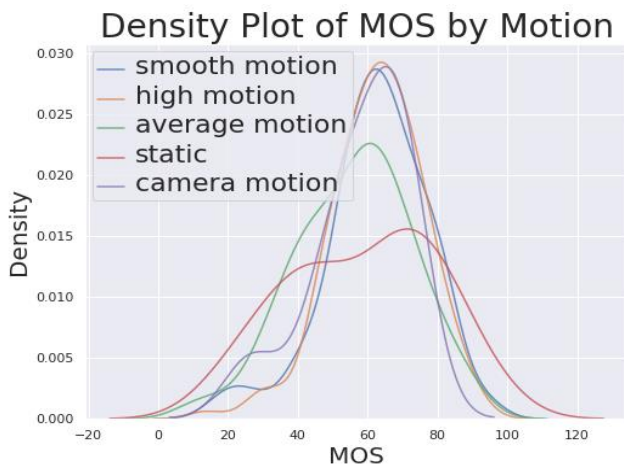

Fig.5. Density plot of MOS by Motion.

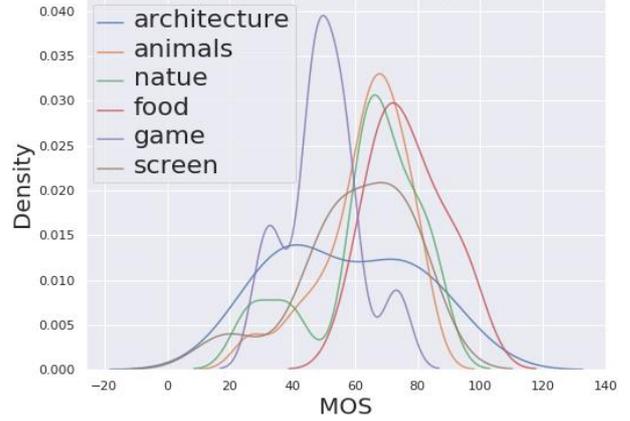

Fig 6. Density plot of MOS by Content types.

Content types differs significantly. However, sports, movie, nature and animation types are similar, human one differs. Animals and nature are similar too, others (screen, food, game and architecture) are differ.

### IV. METHODS AND EXPERIMENTS

Thus, 27 characteristics are obtained for analysis, including 3 categorical ones. We apply machine learning methods (Figure 8) to them. For this, Python 3.7 and sklearn library are used. As a loss function, the Mean Absolute Error will be used to reduce the impact of outliers. To work with categorical features, One-Hot Encoding is used.

### A. QoE curve shape

According to ITU studies [8] QoE dependence on integral quality is sigmoidal:

$$QoE = b_1 / \left[ 1 + exp(b_2(V - b_3)) \right], \quad (1)$$

where b1 is the scaling coefficient along the vertical axis, b2 is slope coefficient and b3 is horizontal offset, $V = \sum_{i=1}^{n} w_i \psi_i$, $i = 1, \ldots, n$, $w_i$ – weight coefficients, $\psi_i$ – objective features. V is the integral quality assessment, some linear or polynomial function. The researched MOS scale is 0-100, so b1 equals 100. The remaining b1 and b2 coefficients are entered in V.

The intuition of sigmoidal behavior lies in the fact that when the quality deteriorates below a certain integral indicator $V_{low}$, the quality assessment does not drop anymore – there is nowhere worse, and when a certain $V_{high}$ is exceeded, the quality assessment does not increase anymore,

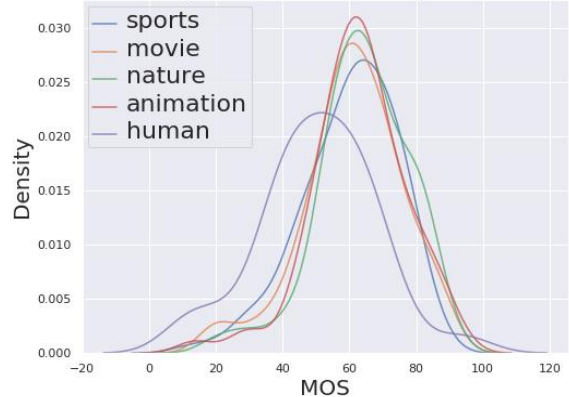

Fig. 7. Density plot of MOS by Content types part 2. It shown that human content type classification is not very reliable feature to analysis.

because the user is completely satisfied.

Thus, we can carry out the inverse transformation and apply it to MOS values.

$$V = \log\left[QoE/(100-V)\right], \quad (2)$$

For methods that use trees (Random Forest, Gradient Boosting and ect.), this is not important (however, restoration of a simpler function is faster and with the help of fewer trees), but for Linear Regression, which we need to create easy-to-use VQA model variants, this fundamentally.

*B. Data preprocessing*

First, we divide the dataset into test, train and validation parts using the stochastic stratified partitioning. This is a common practice for classification tasks, but in regression problems such an approach is rare.

Let *N* denote the number of samples, *y* the target variable for the samples, and k the number of equally sized partitions we wish to create.

With sorted stratification, we first sort the samples based on their target variable *y*. Then we step through each consecutive *k* samples in this order and randomly allocate exactly one of them to one of the partitions. We continue this *floor(N/k)* times, and the remaining *mod(N,k)* samples are randomly allocated to one of the k partitions.

This algorithm ensures the distribution over y is the same in each of the *k* partitions as it is for the *N* samples overall.

Using this algorithm, we create 80/10/10 subsamples parts for train, test and validate. After that data was normalized to the scale [0,1].

Note, that this is very effective instrument (Figure 8), which allowed us to get rid of the overfitting (previously we used the standard ShuffleSplit function from sklearn).

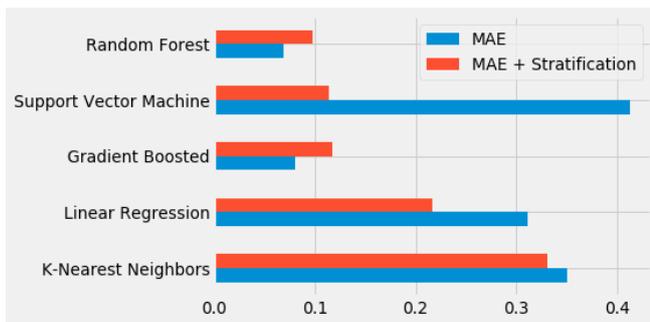

Fig. 8. Stratification applyig to the five basic models with default hyperparametrs.

*C. Basic Machine Learning Models*

A comparison of 5 machine learning models will be made:

- Linear Regression
- Gradient Boosting Regression
- Support Vector Machine Regression
- Random Forest Regression
- K-Nearest Neighbors Regression

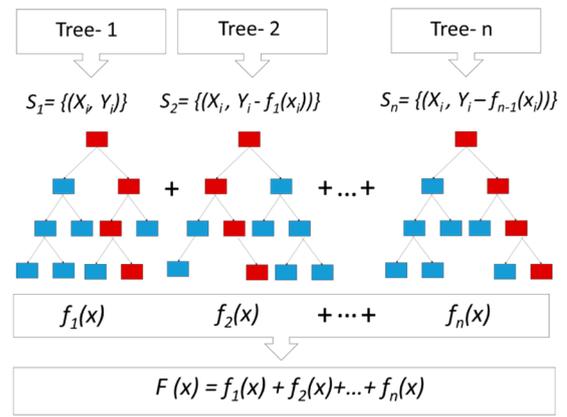

Fig. 9. Schematic description of the work of gradient boosting. Note that a weighting coefficient can be put in front of each $f_i(x)$ and optimized by numerical optimization of the error. Usually, a small value constant is chosen instead, which we did to speed up the calculations..

According to the recommendations of Andrew Ng [31], it is necessary to determine the baseline performance. It can be median value on the test set, Baseline MAE = 0,5554.

*D. Main QoE Model*

Gradient Boosting Regression was chosen as the main model for the study of QoE, as a more powerful tool than the Random Forest. Support Vector Machine was not chosen, because its quality, as shown in Figure 6, is highly dependent on the partition.

Gradient Boosting Regression is an ensemble of simple estimators (in our case trees) that are trained on the residuals of the algorithm of the previous tree. The final algorithm is the weighted sum of the algorithms of individual estimators. The intuition how it works is shown at the Figure 9. The algorithm is described in detail by its authors in their remarkable work [32].

A series of selection of optimal hyperparameters was carried out, at first using a random search on the grid by RandomizedSearchCV method, including ShuffleSplit cross validation using 5 random folds, then the parameters were refined by brut force search by GridSearchCV, also with cross validation (Figure 10).

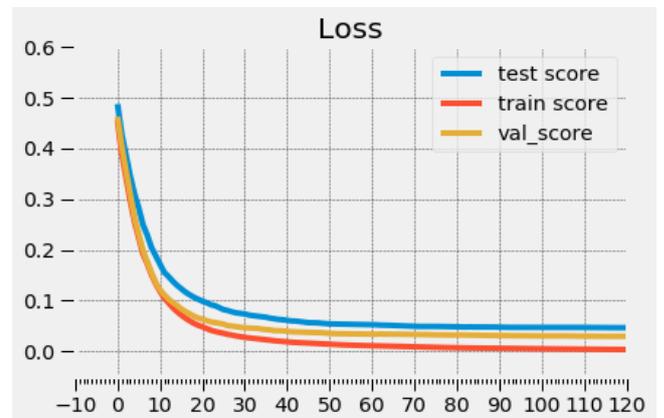

Fig.10. Learning curves on the best hyperparameters. This kind of curves shows that there is no overfitting. The model can be overfitted for both training and test data, but the validation sample is not used when choosing a model and its parametrs. Only at the final assessment.

Using Test set 98 trees was choosen, and the best (and quite simple) model has such hyperparameters:

- max_depth=3
- n_estimators=200
- min_samples_split=10
- min_samples_leaf=6
- loss='huber'
- criterion='friedman_mse'
- max_features='sqrt

The density plot and the distribution of residuals are shown at the Figures 11,12,13. Final model performance is:

- MAE on test set = 0.1031,
- MAE on validation set = 0.0908
- SRCC on test set = 0.9642, P-value = 5.59e-26
- SRCC on validation set = 0.9647, P-value = 1.52e-26

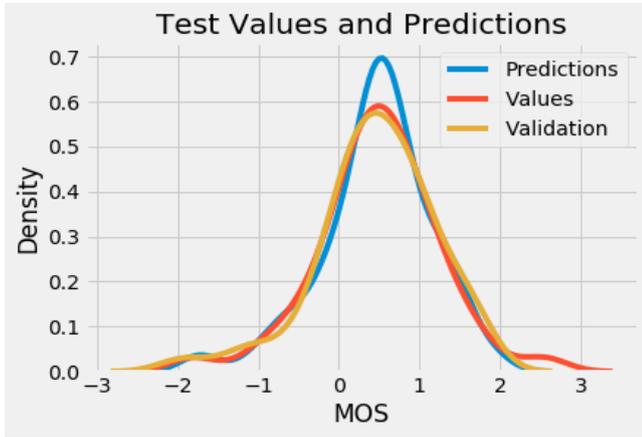

Fig.11. Density plot of Gradient Boosting Regression QoE evaluation.

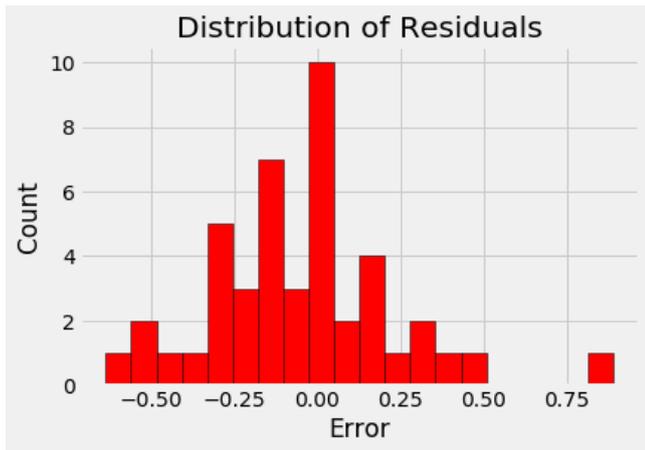

Fig. 12. Distribution of residuals of Gradient Boosting Regression QoE evaluation.

## V. MODEL INTERPRETATION

In this section, the normalization and interpretation of the function of QoE as sigmoid for clarity was removed. The model below is newly trained, it required 597 trees to achieve the same quality.

### A. Feature Importances

Feature importance can be interpreted as the variables,

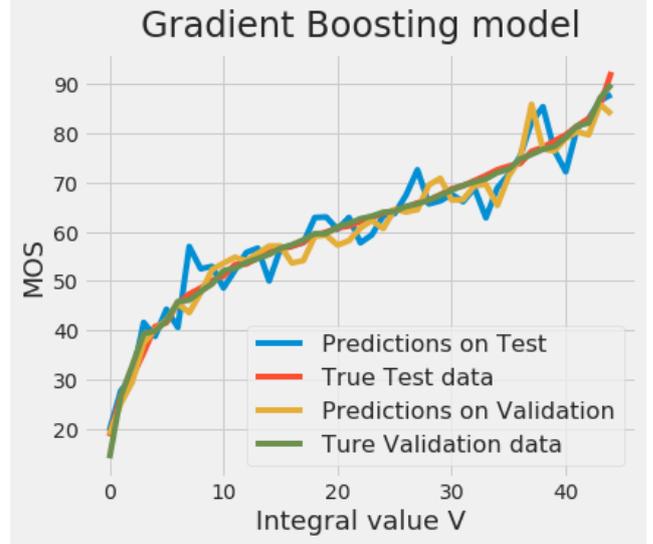

Fig. 13 Gradient Boosting VQA model results. Blue line describers predictions on test set and orange line shows results on validation set.

which are most predictive of the target. To be correct, it is mean decrease impurity and defined as the total decrease in node impurity, weighted by the probability of reaching that node, average all over the trees of the ensemble. Probability of reaching given node approximated by the proportion of samples reaching that node [33]. 10 most important features of Gradient Boosting model is shown in Table IV.

TABLE IV. 10 FEATURE IMPORTANCES OF THE GRADIENT BOOSTED MODEL

| Feature | Importance | Weight |
|---|---|---|
| w0 | | 37.72 |
| Ratio on sequence quality level max/2 | 0.22218 | 17.7497 |
| Average video resolution | 0.1382 | 0 |
| Mean(seqPSNR) | 0.0941 | 0.4884 |
| Ratio on minimum sequence quality level | 0.0930 | -21.7635 |
| Average Weighted Bitrate | 0.0731 | 0.0006 |
| Rebuffer count | 0.0350 | -3.1143 |
| Initial buffer time | 0.0345 | -0.1277 |
| Rebuffer percentage | 0.0293 | -8.9932 |
| Frequency of switching | 0.0286 | -0.0848 |
| Maximum duration of stalling | 0.0278 | -1.4061 |

Using these characteristics as the most significant, a linear regression model was trained:

- MAE on test set = 5,7252,
- MAE on validation set = 5,6419
- SRCC on test set = 0.8436, P-value = 3.5e-13
- SRCC on validation set = 0,8631, P-value = 2.4e-14

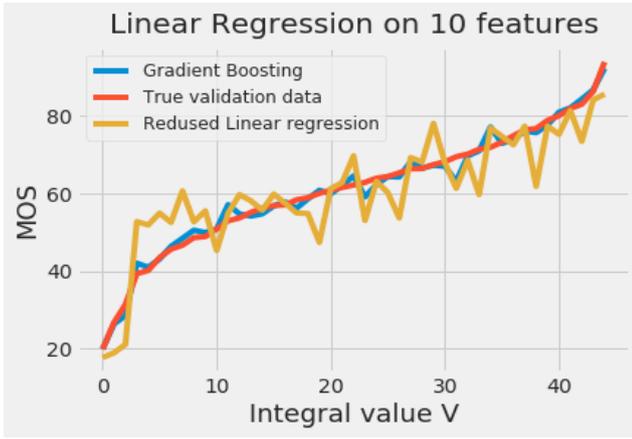

Fig. 14. The Linear model results on the 10 most important signs of the Gradient Boosting model

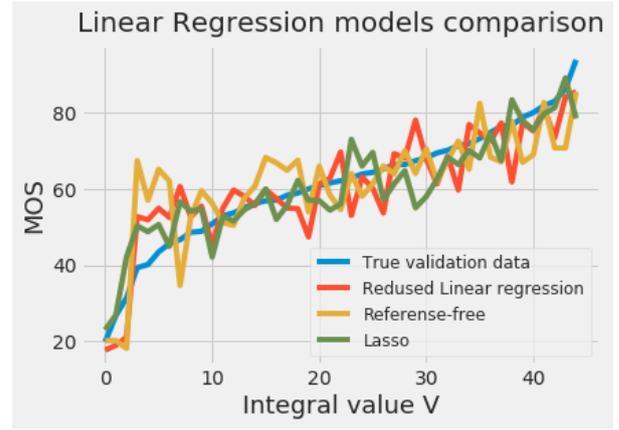

Fig. 15. Comparison of the results of the work of three regression models. The validation subsample is presented as a reference.

Note that this value gives without normalization and applying (2). Basic Linear Regression on full database gives 4.7483, and 4.7902 MAE respectively. Thus, we obtain the formula VQA (Table IV). Comparison with Gradient Boosting model is shown on Figure 14.

Using Linear Regression with Lasso (L1) regularization, the second VQA simple formula was obtained (Table V). L2 regularization and ElasticNet (linear combination of L1 and L2 penalty) gave a little more deadly results. In addition, L1 allows selecting the most important features and simplifying the model. This data is also without normalization, but with knowledge about sigmoid nature of QoE, so target values were transformed by formula (2) and this formula describer value V in formula (1) Parameter of regularization alpha equals 0.00005.

- MAE on test set = 0.2013,
- MAE on validation set = 0.2053
- SRCC on test set = 0.92289, P-value = 2e-19
- SRCC on validation set = 0.9422, P-value = 4.9e-22

TABLE V. DESCRIPTION OF LASSO REGRESSION VQA MODEL

| Feature | Weight | Feature | Weight |
|---|---|---|---|
| w0 | 0.11 | Ratio on sequence quality level max/2 | 0.9112 |
| Rebuffer count | -0.4618 | Average video resolution | 0.3147 |
| Mean(seqPSNR) | 0.3957 | TI | -0.1727 |
| Average Weighted Bitrate | 0.4165 | Content_animals | 0.0639 |
| Maximum duration of stalling | -0.1122 | Content_animation | 0.0923 |
| Bitrate_pos_changes_count | -0.1494 | Content_food | 0.6206 |
| Bitrate_max_pos_change | -0.013 | Content_game | -0.0142 |
| Frequency_of_stalling | -0.0971 | Content_human | -0.0661 |
| Rebuffer percentage | -0.5905 | Content_movie | -0.0956 |
| Ratio on highest sequence quality level | -0.0611 | Motion_average motion | -0.3295 |
| Ratio on minimum sequence quality level | -0.8315 | Motion_smooth motion | 0.0223 |

Zero coefficients for the types of content and motion confirm their dramatically rude rating, that shown above. Nevertheless, non-zero weights for some types of content and movement can be considered confirmation of the importance of developing a more accurate tool for evaluating them. Zero weights with other signs correspond to their low SRCC or/and P-value, but not in every case. For example, constant bitrate is not important, but TI future has enough big weight.

Using the information about the P-value, removing the dubious attribute "Ratio on sequence quality level max/2", motion and content type, removing the need to have a reference video – PSNR knowledge, the simple reference-free model was obtained (Table VI). Lasso Regression was used, alpha = 0.0002. This data has no regularization, but MOS values was changed to sigmoid behavior using formula (2). So, to apply this formula use (1).

- MAE on test set = 0.3421,
- MAE on validation set = 0.327
- SRCC on test set = 0.9054, P-value = 1.3e-17
- SRCC on validation set = 0.9277, P-value = 5.2e-20

TABLE VI. DESCRIPTION OF LASSO REGRESSION VQA REFERENSE-FREE MODEL

| Feature | Weight | Feature | Weight |
|---|---|---|---|
| w0 | 0,31 | Bitrate switch count | 0.1213 |
| Rebuffer count | -0.2369 | Frequency of switching | -0.7385 |
| Average duration of stalling events | -0.0149 | Rebuffer percentage | -1.9522 |
| Average Weighted Bitrate | 0.0001 | Average Bitrate Swithcing magnitude | 0.0001 |
| Maximum duration of stalling | -0.0076 | Average relative Bitrate Swithcing magnitude | -0.0002 |
| Bitrate_neg_changes_count | -0.0105 | Ratio on highest sequence quality level | -0.1628 |
| Bitrate_mean_neg_change | 0.0002 | Ratio on minimum sequence quality level | -1.1528 |
| Frequency of stalling | 1.4992 | Constant_bitrate | 0.1442 |

The Figure 15 shows the results of the Gradient Boosting and Linear Regression on 10 basic characteristics of the model and Reference-free models on validation set comparison with true validation MOS values.

As shown at Figures 13-15, the MOS function is similar to a sigmoid, but rotated 90 degrees and mirrored.

Moreover, the function is limited by the MOS rank scale. It can be assumed that a more accurate description is a composite function of the following form.

$$QoE = \min(scale\_range), V < V_{low}$$
$$QoE = -\log\left(b1 * (b2-V)/V\right) + b3, V \in [V_{low}, V_{high}]$$
$$QoE = \max(scale\_range), V > V_{high}$$

In contrast to ITU research, experimental data show that, starting from the threshold $V_{high}$, quality begins to grow faster than linearly to the maximum, and, accordingly, at a level below the threshold $V_{low}$, it likewise rapidly falls.

### B. Examining a Single Decision Tree

An important advantage of trees-based feasibility to see how each estimator works. The format of the article does not imply the study of each tree of the model, but it is possible to get a general idea of how this VQA model makes decisions. Figure 15 shows the three number 16.

Each node has four pieces of information:

1. Decision rule with a threshold
2. The measure of the error for all of the examples in a given node (Friedman MSE)
3. The number of examples in a given node (samples)
4. The prediction of the target for all examples in a node (value)

Note, that values in our case are residuals. The deeper the leaf in the tree, the less signs in it, the more accurate values it gives. At the same time, deep trees are a risk of overfitting.

For visualization used sklearn.tree.export_graphviz.

## VI. COMPARISON AND RESULTS

Table VII shows the results of various metrics on the dataset. The last value is the result of a regression model by the author of the SQoE-III dataset, the latter is the result of this work.

TABLE VII. RESULTS COMPARISON

| QoE model | SRCC | QoE model | SRCC |
|---|---|---|---|
| SSIMplus [34] | 0,5617 | Mok [42] | 0.1702 |
| VQM [35] | 0,5650 | VsQM [43] | 0.2010 |
| Liu [36] | 0,5145 | Xue [44] | 0.3840 |
| Yun [37] | 0,7143 | Liu [45] | 0.8039 |
| FTW [38] | 0,2745 | SQI [46] | 0.7707 |
| Bentaleb [39] | 0,6322 | P.NATS [47] | **0.8454** |
| ATLAS [40] | 0,1941 | Original Regression model of SQoE-III authors [7] | 0.7800 |
| Kim [41] | 0,0196 | This paper results | **0.9647** |

The aim of this work was to investigate the dependence of various factors on the QoE for the construction of Video Quality Assessment models.

The behavior of the MOS curve was studied, the hypothesis of a sigmoidal dependence on the integral quality indicator on the SQoE-III database was not confirmed, and however, nonlinear sigmoidal transformation of MOS values improves the quality of simple linear models. A composite new formula with a logarithmic dependence was proposed.

The influence of classical and handcrafted features was investigated and shown that:

1) Ratio on lowest and highest quality (in term of bitrate) in sequence has significant impact on MOS (up to -0. 4895)

2) An excellent (SRCC 0.7586) factor is the proportion of playback at half the maximum possible bitrate and higher, but clarification is required, since this factor depends on the distribution of bitrate levels.

The influence of motion estimation in video and the type of content was investigated and it was shown that

1) The human assessment shows almost the same results on a number of types of content and movement

2) Although the signs are not entirely valid, they have a noticeable effect.

The model with SRCC 0.9647 based on Gradient Boosting was created and explained how it works. Besides, several less accurate (SRCC up to 0.9277), but simple to use linear models, including reference-free one, were proposed.

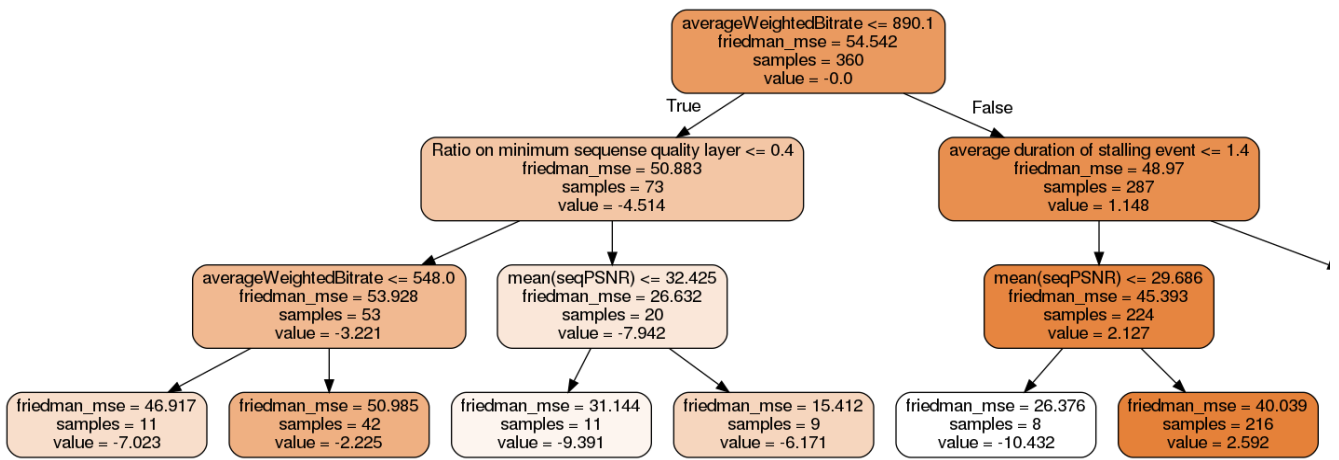

Fig. 16. One of estimators in the Gradien Boosting VQA model (left part).

## DISCUSSION

The study has confirmed that the amount of movement must be measured as a real number. In addition, it is a dynamic indicator. In the initial experiment, this parameter characterizes 10 seconds of the initial video using TI and human qualitative score, but on practice a continuous estimation algorithm is needed. Initial data can be obtained at the codec level from a map of motion vectors.

Estimation of the content type seems non-trivial. On the one hand, it would be possible to create a large database with various types of videos, but its completeness and correctness would always be subject to criticism – there is hardly an answer to the question "what types of content are there?". We propose relying not on classification, but on clustering – there is a hypothesis that not the fact of the type of content itself is important, but its spatial properties, which can be estimated using DNN. Without such a tool, the type of content is not a valid feature.

An important feature is the effect of short-term and long-term memory [50]-[51], which can be estimated using RNN networks. The very first and recent events in a multimedia session can have a more significant impact on QoE.

It is also required to conduct additional research of expert evaluation in the case of very good and very poor quality to restore function at the edges of the distribution. According to the experimental data, the lower limit of the MOS has not dropped below 11/100, which can be both a shift in the beginning of the ranking scale and insufficiently poor video quality.

During conduction of the experiment to expand the database, it will be necessary to determine the typical scenarios of the network and configure them correctly. In the original paper [7], this point was described without details; only the standard set of Linux utilities was specified traffic control and qdisc. In our other study [48], we encountered incorrect traffic control behavior when introducing packet loss - with 10% packet loss set, we had instantaneous burst losses of up to 90%.

Some studies say that it is necessary to increase the duration of the session from 10 to 30 seconds [49]. Moreover, the dataset should be enriched with multimedia sessions of various durations. In addition, of course, it is necessary to expand the base of reference videos both in terms of number and in SI-TI space and the size of the expert group.

In a further study, these and other improvements are planned.

## ACKNOWLEDGMENT

Authors thank the collective of the Laboratory of Multimedia Systems and Technology (LabMST) of Moscow Institute of Physics and Technology for help in this study.